\documentclass[12pt]{article}
\usepackage{graphicx,color,amssymb,amsmath}
\begin{document}
\title{\textbf{Approaching quantum behavior \\with classical fields}} 

\author{B. Holdom%
\thanks{bob.holdom@utoronto.ca%
}\\
\emph{\small Department of Physics, University of Toronto}\\[-1ex]
\emph{\small Toronto ON Canada M5S1A7}}
\date{}
\maketitle
\begin{abstract}
By averaging over an ensemble of field configurations, a classical field theory can display many of the characteristics of quantum field theory, including Lorentz invariance, a loop expansion, and renormalization effects. There is additional freedom in how the ensemble is chosen. When the field mode amplitudes have a Gaussian distribution, and the mode phases are randomly distributed, we review the known differences between the classical and quantum theories. When the mode amplitudes are fixed, or have a nongaussian distribution, the quartic and higher correlations among the free fields are modified, seemingly in a nonlocal way. We show how this in turn affects the perturbative expansion. We focus on  $\lambda\phi^4$ theory in $1+1$ dimensions and use lattice simulations to augment our study. We give examples of how these nonlocal correlations induce behavior more similar to quantum field theory, at both weak and strong coupling.
\end{abstract}

\section{Classical fields masquerading as quantum fields\label{S2}}
We begin by considering free field configurations of a classical field theory having the same energy spectrum as the vacuum fluctuations of quantum field theory. From an ensemble of such configurations the $n$-point correlation functions can be obtained, and we will discuss some dependence that these functions have on the choice of ensemble. The relation to quantum field theory will become more clear in later sections, where we discuss how the free configurations are to be perturbatively corrected to account for interactions. The choice of ensemble that we discuss here will have a nontrivial effect on this perturbative expansion.

For a classical field theory with Lagrangian
\begin{equation}
{\cal L}=\frac{1}{2}\partial_\mu\phi\partial^\mu\phi-\frac{1}{2}m^2\phi^2-\frac{1}{4}\lambda\phi^4
,\label{e2}\end{equation}
we may consider a field configuration consisting of a sum over the free modes,
\begin{equation}
\phi_0(x)=\sum_\mathbf{p}a_\mathbf{p}\cos(\omega_\mathbf{p} t+\mathbf{p}\cdot\mathbf{x}+\theta_{\mathbf{p}})
.\end{equation}
These are solutions for  $\lambda=0$ and $\omega_\mathbf{p}=\sqrt{\mathbf{p}^{2}+m^{2}}$. We will typically work in a finite volume $V$ and so have a discrete set of modes. We constrain the value of the free Hamiltonian for these configurations to be
\begin{equation}
H_0=\sum_\mathbf{p}\frac{1}{2}\hbar\omega_\mathbf{p}
\label{e7}\end{equation}
and thus fix the amplitudes as
\begin{equation}
a_\mathbf{p}^2=\frac{\hbar}{V\omega_\mathbf{p}}
.\label{e10}\end{equation}

Now consider an ensemble of such configurations, differing only through the choice of random phases $\theta_{\mathbf{p}}$ which we take to be uniformly distributed. In the limit of a sufficiently large number of configurations we can obtain the expectation value of a product of fields by simply averaging over the phases. This gives \cite{F,C}
\begin{equation}
\langle\phi_{0}(x)\phi_{0}(y)\rangle\equiv D^0(x-y)=\sum_\mathbf{p}\frac{\hbar}{2\omega_\mathbf{p} V}\cos(p(x-y)) \mbox{,  with $p=(\omega_\mathbf{p},\mathbf{p})$}
\label{e14}.\end{equation}
In the $V\rightarrow\infty$ limit we could write this as
\begin{eqnarray}
\langle \phi_0(x)\phi_0(y)\rangle&=&\int \frac{d^4p}{(2\pi)^4} G_{\phi\phi}^0(p)e^{-ip(x-y)},\\
G_{\phi\phi}^0(p)&=&\hbar\pi\delta(p^2-m^2)
,\label{e9}\end{eqnarray}
thus showing that the $\frac{1}{2}\hbar\omega$ spectrum is Lorentz invariant in this limit.

A more conventional choice of ensemble would have the mode amplitudes also varying from one configuration to the next, in such a way that only the average value of the energy of each mode is 
$\frac{1}{2}\hbar\omega_\mathbf{p}$. Then the amplitudes can have a Gaussian probability distribution,
\begin{equation}
\rho(a_\mathbf{p})=\frac{2 a_\mathbf{p}}{\sigma_\mathbf{p}^2}\exp(- \frac{a_\mathbf{p}^2}{\sigma_\mathbf{p}^2}),\quad\quad\sigma_\mathbf{p}^2=\hbar/(V\omega_\mathbf{p})
,\label{e8}\end{equation}
while treating the $a_\mathbf{p}$ as nonnegative polar variables. The result (\ref{e14}) for the 2-point function again follows. This choice bears a resemblance to the state vectors of a quantum field theory when written as wave functionals $\Omega[\phi]$, since the solution to the free functional Schrodinger equation is of course a Gaussian.\footnote{Gaussian distributions also arise for classical fields in thermal equilibrium, but the ensembles we are considering are very far from thermal.}

The choice of ensemble affects the quartic and higher free $n$-point functions. They are obtained as before by taking the ensemble average of products of fields, which in turn are sums of modes. For example the free 4-point function in the Gaussian case is as expected,
\begin{eqnarray}
&&\langle\phi_{0}(x_{1})\phi_{0}(x_2)\phi_{0}(x_3)\phi_{0}(x_{4})\rangle_\mathrm{G} =\nonumber\\&&\;\;\;\;D^0(x_{1}-x_{2})D^0(x_{3}-x_{4})+D^0(x_{1}-x_{3})D^0(x_{2}-x_{4})+D^0(x_{1}-x_{4})D^0(x_{2}-x_{3}).\label{e16}\end{eqnarray}
When the amplitudes do not vary between configurations, the fixed amplitude case, there are ``unusual'' contributions in the 4-point function due to terms where the product of modes involves only a single mode, so the four factors share a single phase and a single amplitude. The average is now over the phase only, and this changes the result to
\begin{eqnarray}
&&\langle\phi_{0}(x_{1})\phi_{0}(x_2)\phi_{0}(x_3)\phi_{0}(x_{4})\rangle_\theta =\langle\phi_{0}(x_{1})\phi_{0}(x_2)\phi_{0}(x_3)\phi_{0}(x_{4})\rangle_\mathrm{G}\nonumber\\&&\;\;\;\;\;-\;\frac{1}{2}\Delta(x_1,x_2;x_3,x_4)-\frac{1}{2}\Delta(x_1,x_3;x_2,x_4)-\frac{1}{2}\Delta(x_1,x_4;x_2,x_3),\\
&&\Delta(x_1,x_2;x_3,x_4)=\sum_\mathbf{p}\frac{1}{(2\omega_\mathbf{p} V)^2}\cos(p(x_1-x_2))\cos(p(x_3-x_4)).\label{e12}\end{eqnarray}
Alternatively we may write this as
\begin{eqnarray}
&&\langle\phi_{0}(x_{1})\phi_{0}(x_2)\phi_{0}(x_3)\phi_{0}(x_{4})\rangle_\theta =\nonumber\\&&\;\;\;\;[[D^0(x_{1}-x_{2})D^0(x_{3}-x_{4})+D^0(x_{1}-x_{3})D^0(x_{2}-x_{4})+D^0(x_{1}-x_{4})D^0(x_{2}-x_{3})]]_\frac{1}{2}.\label{e11}\end{eqnarray}
$[[..]]_\frac{1}{2}$ indicates that we must reduce the terms with a common $\mathbf{p}$, after inserting (\ref{e14}) for $D^0$, by a factor of $1/2$. 

In the case of the free 6-point function, contributions with two of the momenta the same are again reduced by $1/2$, while the terms with all three of the momentum the same are reduced by a factor of $1/6$. More generally, contributions to free correlation functions with $n$ momenta the same are reduced by a factor of $1/n!$.

Of course one could engineer a probability distribution $\rho(a_\mathbf{p})$ to produce quite arbitrary changes in the coincident momentum behavior of free correlation functions. In terms of the second moment of the distribution, $\sigma_\mathbf{p}^2$, if the $(2n)$th moment for $n>1$ is $C_n\sigma_\mathbf{p}^{2n}$, then the contributions with $n$ momenta coincident at $\mathbf{p}$ are changed by a factor of $C_n/n!$. The Gaussian and fixed amplitude cases have $C_n=n!$ and $C_n=1$ respectively, and it can be seen that the fixed amplitude case minimizes the $C_n$. A simple family of distributions is
\begin{equation}
\rho(a_\mathbf{p})\propto \frac{a_\mathbf{p}^{d-1}}{\sigma_\mathbf{p}^d}\exp(- \frac{d a_\mathbf{p}^2}{2\sigma_\mathbf{p}^2}),\quad0<d<\infty
\end{equation}
where Gaussian and fixed amplitude cases have $d=2$ and $d\rightarrow\infty$ respectively.

Thus we see that the freedom in the choice of ensemble modifies the free higher $n$-point functions, while holding the free $2$-point function fixed. We notice that $\Delta$ in (\ref{e12}) does not factorize and it does not vanish for large spacelike separations of pairs of points. These modifications then violate cluster decomposition and thus lie outside the realm of correlations produced by a local quantum field theory. In this sense they could be considered nonlocal, even though we see that they can occur naturally in the free field configurations of a local classical field theory. We will refer to them as nonlocal for lack of a better description. As a measure of the nonlocality in the 4-point function we have
\begin{equation}
\beta\equiv2-C_2=\frac{d-2}{d}
,\label{e19}\end{equation}
so that $[[..]]_\frac{1}{2}$ in (\ref{e11}) is replaced by $[[..]]_{1-\beta/2}$.

In the classical field theory these free correlations form the basis of the perturbative expansion, and thus the first question is whether this expansion can be significantly affected by the nonlocal correlations. The perturbative expansion is a loop expansion involving momentum loop integrations, and it might appear that the nonlocal correlations would have vanishing effect in the infinite volume limit since the correction terms involve fewer momentum sums. On the other hand the loop integrands may display singular behavior in regions of momentum space where these corrections contribute. We will show, both through a direct sample calculation and by lattice simulation, that the nonlocal correlations do in fact alter the perturbative expansion in significant ways.

The Gaussian classical theory is known to deviate from the behavior of the corresponding interacting quantum field theory, and we will remind the reader of the differences at the perturbative level. In certain situations of high temperature and/or high occupation number, these differences may be controlled and the classical theory used as an approximation scheme  \cite{G}. But at zero temperature the departures from the quantum theory that are evident for the Gaussian classical theory could be taken to define what truly quantum phenomena are. Thus it is of interest to compare these truly nonclassical effects of the quantum theory to the nonlocal effects of the nongaussian classical theory. We will find that the nonlocal correlations of the free fields can act to bring the perturbative expansion of the classical theory closer to that of the quantum theory. This helps to explain the findings of \cite{C,D}. 

In the next section we will develop the Gaussian classical theory, highlighting both its resemblance to quantum field theory and its qualitative differences. Section 3 will pinpoint the perturbative difference, thus making clear what is missing relative to the quantum theory. In section 4 we consider a 2-loop example where we identify the missing piece, and we compare it to a similar contribution coming from the nonlocal correlations in the nongaussian classical theory. In sections 5 and 6 we use lattice simulations of the classical theory to investigate this more fully, and we find further evidence of quantum-like perturbative effects. In particular damping and thermalization effects are greatly reduced. The simulations can also be used at strong coupling, and in section 7 we will see quantum-like critical behavior emerging in the nongaussian classical theory. We end with some comments and open questions in the final section.

\section{The Gaussian classical theory}

It is possible to develop a perturbative expansion by writing the field equation as an integral equation and then iterating \cite{C}. Instead we shall make use of a path integral that does nothing but enforce the classical field equations, and then obtain the perturbative expansion from the path integral in the usual way. The integration over the fields in the path integral will produce the sum over the ensemble of classical field configurations.

The expectation values of products of fields in some ensemble of configurations can be expressed as follows,
\begin{equation}
\langle \phi(x_1)\phi(x_2)...\rangle=\int d\phi d\chi\; \rho[\phi(x_i)]\phi(x_1)\phi(x_2)...e^{\frac{i}{\hbar}\int_{t_i}^{t_f}d^4x {\cal L}_\chi[\phi,\chi]}
\label{e5}\end{equation}
where $t_i<\{t_1,t_2,...\}<t_f$. (A similar construction appears for example in \cite{ms}.) The Lagrangian ${\cal L}_\chi$ is linear in the auxiliary field $\chi$
\begin{equation}
{\cal L}_\chi=\partial_\mu\phi\partial^\mu\chi-m^2\phi\chi-\lambda\phi^3\chi
,\label{e17}\end{equation}
and we may require that $\chi(x)$ vanishes at times $t_i$ and $t_f$. The functional integral over $\chi$ gives the constraint
\begin{equation}
\prod_{\tau,\mathbf{x}}2\pi\hbar\delta\left[(\Box+m^2)\phi(\tau,\mathbf{x})+\lambda\phi^3(\tau,\mathbf{x})\right]
\end{equation}
for $t_i<\tau<t_f$. In this way the field $\phi$ is constrained to satisfy the interacting classical field equation. Notice that a $\hbar^{-1}$ has been inserted in the action, even though it has no significance at this stage.

$\rho[\phi(x_i)]$, where $x_i$ is the coordinate on the initial time-slice $t=t_i$, selects the configurations in our ensemble. Since we are interested in a perturbation theory it is sufficient to describe this ensemble in terms of the free field modes, and in this section we are making the Gaussian choice. These configurations are acting as the initial conditions for evolution by the full field equations.

Now the point of this construction is to leave the $\chi$ field in the theory to develop the perturbation theory.  The Lagrangian specifies the $\phi\chi$ and $\chi\phi$ propagators up to boundary conditions, and since we are studying a classical theory it is appropriate to choose the retarded and advanced Greens functions,
\begin{equation}
G_{\phi\chi}^0=\frac{i\hbar}{p^2-m^2+i\varepsilon p_0}\;\;\;\;\;G_{\chi\phi}^0=\frac{i\hbar}{p^2-m^2-i\varepsilon p_0}
.\end{equation}
With this choice we see diagrammatically that the $\chi$ end of a line always appears at a later time than the $\phi$ end. But each vertex has only one $\chi$ and three $\phi$'s, so closed loops of $G_{\phi\chi}^0$  or $G_{\chi\phi}^0$ lines do not form. The resulting tree graph expansion is the expected result for a classical theory. 

This would be all in the absence of background fields, but due to our ensemble of field configurations we also have $G_{\phi\phi}^0(p)=\hbar\pi\delta(p^2-m^2)$.  The $\hbar$ appearing here, originating in the mode amplitudes, does have significance. Now the branches of the various trees can be interconnected with $G_{\phi\phi}^0$ lines, and a nontrivial loop expansion emerges. In fact the same relation between the number of loops in a diagram and the power of $\hbar$ occurs in this classical theory as it does in quantum field theory. The $\hbar^{-1}$ in the action was not necessary for this but it makes this result easier to see, since then factors of $\hbar$ are associated with propagators and vertices in the usual way.

The free 2-point functions are also such that $G_{\chi\chi}^0=0$ and
\begin{equation}
G_{\phi\phi}^0=\frac{1}{2}\epsilon(p_0)(G_{\phi\chi}^0-G_{\chi\phi}^0)
.\label{e1}\end{equation}
We can make some statements about the full 2-point functions (denoted by $G$) and the 2PI self-energies (denoted by $\Pi$). Diagramatically it can be checked that $\Pi_{\phi\phi}=0$, and similarly $G_{\chi\chi}=0$ continues to hold in the interacting theory. Then the full $G_{\phi\chi}$ and $G_{\chi\phi}$ satisfy
\begin{eqnarray}
G_{\phi\chi}&=&G^0_{\phi\chi}+G^0_{\phi\chi}\Pi_{\chi\phi}G_{\phi\chi} \\
G_{\chi\phi}&=&G^0_{\chi\phi}+G^0_{\chi\phi}\Pi_{\phi\chi}G_{\chi\phi} 
\end{eqnarray}
and thus
\begin{eqnarray}
G_{\phi\chi} & = & \frac{i\hbar}{p^2-m^2-\Pi_r+i(\Pi_i+\varepsilon p_0)},\;\;\;\;\;\;i\hbar\Pi_{\chi\phi}  \equiv \Pi_r-i\Pi_i  \\
G_{\chi\phi} & = & \frac{i\hbar}{p^2-m^2-\Pi_r-i(\Pi_i+\varepsilon p_0)},\;\;\;\;\;\;i\hbar\Pi_{\phi\chi}  \equiv  \Pi_r+i\Pi_i  
\end{eqnarray}
These $G$'s have the same analytic structure as in the free theory, so $\Pi_i$ is an odd function of $p_0$ with $\Pi_i(p_0)>0$ for $p_0>0$.

Of more interest is the full 2-point function $G_{\phi\phi}$ for the original scalar field. The summation of the diagrams involves various classes of diagrams.
\begin{eqnarray}
G_{\phi\phi}&=&G^0_{\phi\phi}+G^0_{\phi\phi}\Pi_{\phi\chi}G_{\chi\phi}+G_{\phi\chi}\Pi_{\chi\phi}G^0_{\phi\phi}+G_{\phi\chi}\Pi_{\chi\phi}G^0_{\phi\phi}\Pi_{\phi\chi}G_{\chi\phi}+G_{\phi\chi}\Pi_{\chi\chi}G_{\chi\phi}\\
&=&G_{\phi\chi}\Pi_{\chi\chi}G_{\chi\phi}\label{e3}
\end{eqnarray}
The cancellations among the various classes follow from the expressions for $G^0_{\phi\phi}$, $G_{\phi\chi}$ and $G_{\chi\phi}$.  We can compare this to the result of using (\ref{e1}) as a relation among the full 2-point functions, which gives
\begin{eqnarray}
G_{\phi\phi}&=&\frac{\epsilon(p_0)\Pi_i}{(p^2-m^2-\Pi_r)^2+\Pi_i^2}\label{e15}\\
&=&G_{\phi\chi}[\epsilon(p_0)\Pi_i] G_{\chi\phi}\label{e4}
\label{e6}\end{eqnarray}
Comparison of (\ref{e3}) and (\ref{e4}) would imply the relation $\Pi_{\chi\chi}=\epsilon(p_0)\Pi_i$. This is in fact true, as can be checked diagramatically. 

It turns out that $\Pi_i$ (or $\Pi_{\chi\chi}$) is nonvanishing on mass shell, in which case we have the finite width (and shifted mass due to $\Pi_r$) form of a pseudoparticle correlator. The emergence of a pseudoparticle is typical in classical field theory applications, in thermal systems for example, and the pseudoparticle width is often referred to as plasmon or Landau damping. A sampling of references to this and related applications of classical field theories is given in \cite{therm}.

This pseudoparticle distinguishes the classical theory from the particle description of a quantum field theory. Nevertheless there is a loop expansion characterized by powers of $\hbar$, and a renormalization of the parameters $m$ and $\lambda$ in ${\cal L}_\chi$ similar to quantum field theory.  In addition to the $\chi\chi$ amplitude, the classical theory will perturbatively generate amputated amplitudes of the form $\phi^r\chi^s$ with $r+s$ even. We have seen that a $\phi\phi$ self-energy is not generated and more generally amplitudes are not generated unless $s\ge 1$. A real contribution to  the $\phi\chi^3$ amplitude is expected, and in the next section we will find this amplitude to be of special interest.

[Another implication is that ${\cal L}_\chi$ is not corrected by a constant term, a cosmological constant. Here it is useful again to emphasize the difference between the Lagrangian ${\cal L}_ \chi$ appearing in the path integral and the original classical Lagrangian ${\cal L}$ in (\ref{e2}). It is the latter the defines the energy momentum tensor, whose the expectation value can be calculated perturbatively using the path integral involving ${\cal L}_\chi$. A contribution to the cosmological constant could be obtained in this way.]

\section{A small change and the true quantum theory}
Now let us consider a new theory by adding a $\phi\chi^3$ interaction to our previous theory defined by ${\cal L}_\chi$ in (\ref{e5}-\ref{e17}), where we continue to assume the Gaussian ensemble. Denoting the new Lagrangian by ${\cal L}_\alpha$,
\begin{equation}
{\cal L}_\alpha={\cal L}_\chi+\alpha\frac{\lambda}{4}\phi\chi^3
,\end{equation}
it is clear that the direct link to the classical evolution via integrating out the $\chi$ field has been lost. In fact this additional interaction term is all that separates the classical theory from the full quantum field theory. (This observation \cite{G} of the relation between classical and quantum field theories originates in studies of classical fields used to model quantum fields at high temperature or high occupation number.) This may be somewhat surprising, given that we have just observed that this amplitude is generated at some level in the classical theory.

The following redefinition of fields
\begin{equation}
\chi=\sqrt{\frac{1}{\alpha}}(\phi_+-\phi_-)
,\;\;\;\;\;\;
\phi=\frac{1}{2}(\phi_++\phi_-)
,\end{equation}
yields
\begin{equation}
{\cal L}_\alpha=\sqrt{\frac{1}{\alpha}}\left[{\cal L}(\phi_+)-{\cal L}(\phi_-)\right]
,\end{equation}
where ${\cal L}$ is the original classical Lagrangian in (\ref{e2}). Thus ${\cal L}$ is now appearing in the path integral itself. If $\alpha=1$ the overall factor in the action is the usual one for a quantum field theory, given the $\hbar^{-1}$ factor already in the action of (\ref{e5}). 

In this case of $\alpha=1$ the transformed versions of the original free propagators take the following form.
\begin{eqnarray}
G_{++}^0&=&\frac{i\hbar}{p^2-m^2+i\epsilon}\\
G_{--}^0&=&\frac{i\hbar}{p^2-m^2-i\epsilon}\\
G_{-+}^0&=&2\hbar\pi\theta(p_0)\delta(p^2-m^2)\\
G_{+-}^0&=&2\hbar\pi\theta(-p_0)\delta(p^2-m^2)
\end{eqnarray}
$G_{++}^0$ is the Feynman propagator. We have now arrived at the Schwinger-Keldysh formalism for the calculation of ``in-in'' matrix elements in quantum field theory \cite{rj}. This formalism is complementary to the more standard formalism of quantum field theory, concerned with the calculation of ``in-out'' matrix elements.

For example \cite{rj} consider the full $G_{++}$ 2-point function
\begin{equation}
G_{++}(x_1,x_2)\propto\int d\phi_+ d\phi_-\; \rho[\phi_+(x_i)]\phi_+(x_1)\phi_+(x_2)e^{\frac{i}{\hbar}\int_{t_i}^{t_f}d^4x ({\cal L}(\phi_+)-{\cal L}(\phi_-))}
\end{equation}
The original constraints on the $\chi$ field now read $\phi_-(x_i)=\phi_+(x_i)$ and $\phi_-(x_f)=\phi_+(x_f)$. Other than this the $\phi_+$ and $\phi_-$ path integrals have decoupled, and each is a path integral that can be identified with a ``in-out'' transition amplitude or correlation function of quantum field theory. Thus
\begin{eqnarray}
G_{++}&=&\sum_s\langle \textrm{in, vac}|\textrm{out, }s\rangle 
\langle \textrm{out, }s|T\phi(x_1)\phi(x_2)|\textrm{in, vac}\rangle \\
&=&\langle \textrm{in, vac}|T\phi(x_1)\phi(x_2)|\textrm{in, vac}\rangle
\end{eqnarray}
$G_{--}$, $G_{+-}$ and $G_{-+}$ have analogous expressions. The sum over ``out'' states corresponds to the fact that no constraint has been placed on $\phi(x_f)$ in the path integral. And $\rho[\phi(x_i)]$ has given us an explicit representation of what we mean by the ``in'' vacuum state; for the Gaussian choice we have completed the equivalence to quantum field theory.

Thus the modification of the theory through the addition of the $\phi\chi^3$ term has replaced the classical correlation functions by their quantum counterparts. We see that the whole essence of quantum field theory (for a scalar field) lies in a particular value for the coefficient of the $\phi\chi^3$ term. The essence is not the $\hbar$, or the loop expansion, or renormalization, since all this appears in the classical theory. And as the coefficient $\alpha$ varies from 0 to 1, the theory extrapolates continuously from classical to quantum.

\section{Quantum behavior at 2-loops\label{S1}}

In this section we first isolate a truly quantum perturbative contribution to the self-energy, a contribution that is due to the $\phi\chi^3$ term. We will see explicitly that this contribution for $\alpha=1$ has the effect of cancelling the classical contribution to the imaginary self-energy $\Pi_i$ on mass shell. Then the width of the pseudoparticle goes to zero, and we end up with a stable particle as described by quantum field theory. But the main point here will be to continue our discussion of section 1, and obtain the nonlocal corrections to the same diagram that can occur in the classical theory.

\begin{figure}
\begin{center}\includegraphics[%
  scale=0.5]{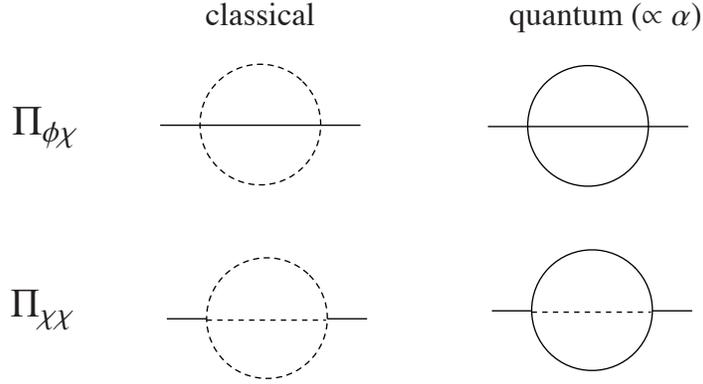}\end{center}
  \vspace{-3ex}
\caption{The 2-loop self-energy diagrams are split into classical and quantum pieces. The solid lines are $G^0_{\phi\chi}$ or $G^0_{\chi\phi}$ and the dashed lines are $G^0_{\phi\phi}$.\label{C}}
\end{figure}

A truly quantum contribution to the self-energy first appears at the 2-loop level. We return to the $(\phi,\chi)$ basis where a $\phi\chi^3$ term generates loops of $G^0_{\phi\chi}$ or $G^0_{\chi\phi}$ lines, thus allowing more than one $G^0_{\phi\chi}$ or $G^0_{\chi\phi}$ line to connect two vertices. We show the resulting classical and quantum pieces of the sunset diagram in Fig.~(\ref{C}). We have seen that the $\Pi_{\chi\chi}$ diagrams are just the imaginary parts of the $\Pi_{\phi\chi}$ ones, so we need only consider the top two diagrams.

We restrict ourselves to $1+1$ dimensions as we do in the lattice simulations to follow, and consider the classical 2-loop contribution to $\Pi_{\phi\chi}$, the first diagram in Fig.~(\ref{C}). Without Wick rotating and in finite volume $V=L$ this is
\begin{equation}
\Pi_\textrm{classical}^\textrm{2-loop}(k)=\frac{6!^2\lambda^2}{2L^2}\sum_{p_1,q_1}\int\frac{dp_0}{2\pi}\frac{dq_0}{2\pi}\frac{\pi\delta(p_0^2-p_1^2-m^2)\pi\delta(q_0^2-q_1^2-m^2)}{(k_0-p_0-q_0+i\epsilon)^2-(k_1-p_1-q_1)^2-m^2}
.\label{e13}\end{equation}
After the temporal momentum integrals are done we are left with sums over the spatial momenta $p_1$ and $q_1$ whose discrete values are multiples of $2\pi/L$. These sums are infrared dominated. They are evaluated numerically while keeping the $\epsilon$ in the retarded propagator small but nonvanishing. Of interest is the on-shell self-energy, obtained by setting $(k_0,k_1)=(m,0)$. In a similar way we may consider the quantum contribution proportional to $\alpha$, the second diagram in Fig.~(\ref{C}).

Combining the results for the real and imaginary parts we have
\begin{eqnarray}
\Pi_r^\textrm{2-loop}(m) & \propto & 1-\alpha/3,  \\
\Pi_i^\textrm{2-loop}(m) & \propto & 1-\alpha.
\end{eqnarray}
Thus in the quantum theory defined by  $\alpha=1$ we find that $\Pi_r^\textrm{2-loop}(m)$ is 2/3 of its size in the classical theory.  And $\Pi_i^\textrm{2-loop}(m)$ vanishes as expected.  The classical and quantum pieces of $\Pi_r^\textrm{2-loop}(m)$ are each finite and quite insensitive to how the $V\rightarrow\infty$ and $\epsilon\rightarrow0$ limits are taken. For $\Pi_i^\textrm{2-loop}(m)$ on the other hand there is a weak divergence in each piece as $\epsilon\rightarrow0$, only canceling when $\alpha=1$. We note that this divergence is an artifact of the perturbative calculation, since the full $\Pi_i(m)$ is certainly finite in the classical theory. We will see this in the simulations. Although we will not pursue it here, one could consider replacing the free propagators by the full propagators $G_{\phi\phi}$ and $G_{\phi\chi}$ in the 2-loop calculation.

Now that we have isolated the ``truly quantum'' from the classical, we may consider the effect of making another choice for the ensemble in the classical theory. We return to the classical theory defined by $\alpha=0$, but now account for a modified perturbative expansion corresponding to the different choice of $\rho[\phi(x_i)]$ appearing in (\ref{e5}). According to section 1, as we depart from the Gaussian prescription the free $n$-point functions are no longer simply expressed in terms of the free $2$-point function. We parametrized the nonlocality in the 4-point function by $\beta$, where $\beta=0$ and $\beta=1$ correspond to the Gaussian and fixed amplitude cases respectively. The question is how a nonvanishing $\beta$ affects the 2-loop calculation in (\ref{e13}). 

The Feynman rules for the two $G_{\phi\phi}^0$ lines in the diagram arose by writing the $D^0$ in (\ref{e14}) as
\begin{equation}
D^0(x-y)=\sum_{p_1}\frac{1}{V}\int \frac{dp_0}{2\pi} G_{\phi\phi}^0(p)e^{-ip(x-y)}
\label{e18}.\end{equation}
We notice that relative to (\ref{e14}) this incorporates some inconsequential changes of sign of $p_1$ in the terms that have negative $p_0$. But to identify which contributions in (\ref{e13}) need to be modified, we should undo these changes of sign by replacing $p_1$ inside the sum in (\ref{e18}) by $\mathrm{sign}(p_0)p_1$, and the same for $p_1$ and $q_1$ in (\ref{e13}). Then the rule from section 1 is to reduce the terms with $p_1=q_1$ in the discrete momentum sums by a factor of $(1-\beta/2)$. The net result is easier to describe in terms of the two 2-vectors $p_\mu$ and $q_\mu$, now with the sign factors included in their definition. The rule is to reduce by $(1-\beta/2)$ the contributions where the 2-momenta of the two $G_{\phi\phi}^0$ lines satisfy either $p_\mu=q_\mu$ or $p_\mu=-q_\mu$.\footnote{We notice the Lorentz invariance of this rule.}

\begin{figure}
\begin{center}\includegraphics[%
  scale=0.6]{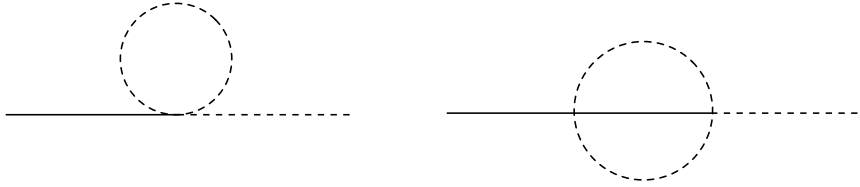}\end{center}
  \vspace{-3ex}
\caption{1-loop and 2-loop contributions to the 2-point function.\label{D} }
\end{figure}

But this is not all, since what is directly measurable is the 2-point function rather than the self-energy. And one of the two external lines of the sunset diagram contribution to the 2-point function must be a $G_{\phi\phi}^0$ line (second diagram in Fig.~(\ref{D})). The same is true at 1-loop (first diagram in Fig.~(\ref{D})), which in this case implies that there are two $G_{\phi\phi}^0$ lines in that diagram. Thus there is also a nonlocal correction to the 1-loop contribution to the 2-point function. For an external line at vanishing spatial momentum, the correction involves the zero momentum contribution to the loop. This then is a factor of $(1-\beta/2)$ smaller than usual while all the other modes contribute in the loop as usual. It can easily be checked that this nonlocal correction at 1-loop will vanish in the $V\rightarrow\infty$ limit for fixed mass.

Returning to the sunset diagram, the external $G_{\phi\phi}^0$ line means that there are three $G_{\phi\phi}^0$ lines in this contribution to the 2-point function. Thus when either of the internal $G_{\phi\phi}^0$ lines has vanishing spatial momentum, these contributions should also be reduced by $(1-\beta/2)$. Unlike the 1-loop case, we find that these corrections, along with the previous $p_\mu=q_\mu$ or $p_\mu=-q_\mu$ corrections, do not vanish in the $V\rightarrow\infty$ limit.\footnote{The contribution when all three spatial momenta vanish does vanish.} They produce a finite correction to $\Pi_r^\textrm{2-loop}(m)$ as long as the limit $\epsilon\rightarrow0$ is taken before the limit $V\rightarrow\infty$ (in the opposite order it vanishes). More precisely $\epsilon$ must be sufficiently small compared to lattice spacing $2\pi/L$ of the momentum space lattice used to evaluate the sums. The finite correction illustrates how the singularity structure of momentum loop integrals, arising from $\epsilon\rightarrow0$, can offset the naive volume suppression of the correction. 

The actual result from numerical analysis is
\begin{eqnarray}
\Pi_r^\textrm{2-loop}(m) & \propto & 1-\beta/2,  \\
\Pi_i^\textrm{2-loop}(m) & \propto & 1-\beta/2.\label{e20}
\end{eqnarray}
Each of the three ways of having pairwise coincident momentum in the three $G_{\phi\phi}^0$  lines contribute equally in $\Pi_r^\textrm{2-loop}(m)$, whereas only the case of coincident momentum on the two internal lines contributes in $\Pi_i^\textrm{2-loop}(m)$. But since each result is proportional to $1-\beta/2$, this shows that \textit{only} the coincident momentum terms contribute; the contributions without coincident momenta vanish!

We see that $\Pi_r^\textrm{2-loop}(m)$ can be reduced from the Gaussian classical value as in the quantum theory. $\Pi_i^\textrm{2-loop}(m)$ is still plagued with the overall $\epsilon\rightarrow0$ divergence, so result (\ref{e20}) is not particularly useful. To see how this quantity is actually behaving we turn to the lattice simulations of the classical theory ($\alpha=0$). We will be comparing the Gaussian $\beta=0$ and the fixed amplitude $\beta=1$ cases.

\section{Lattice simulations}
We wish to obtain both the real and imaginary parts of the self-energy through a direct lattice study of interacting classical fields in $1+1$ dimensions. The details of the simulations are found in \cite{C,D}. They are performed by numerically evolving each configuration according to the full field equations in real time. For lattice spacing $a$, the spatial size of the lattice is $Na$ where $N=256$ unless otherwise specified. The definition of the initial configurations requires a mass parameter, and this is chosen to match the physical mass as determined by the simulation. The expectation values of products of fields are directly obtained in coordinate space. An accurate determination of the mass corrections can be more easily determined from time-like rather than space-like correlators, since the latter falls exponentially. We will consider two time-like correlators, where the first one involves the zero mode only,
\begin{eqnarray}
D_\mathit{zm}(t)&=&\int\frac{dp_0}{2\pi} \left. G_{\phi\phi}(p)\right|_{p_1=0}e^{-i p_0 t},\\
D_t(t)&=&\frac{1}{L}\sum_{p_1}\int\frac{dp_0}{2\pi}G_{\phi\phi}(p)e^{-i p_0 t}
.\end{eqnarray}
In the free case with $G_{\phi\phi}\rightarrow G_{\phi\phi}^0=\hbar\pi\delta(p^2-m^2)$ we have
\begin{equation}
D_\mathit{zm}^0(t)=\frac{1}{2m}\cos(m t),\;\;\;\;\;D_t^0(t)=\frac{1}{L}\sum_{p_1}\frac{1}{2\omega}\cos(\omega t)
.\end{equation}
When we instead use $G_{\phi\phi}$ from (\ref{e15}), and use a narrow width approximation where the self-energy functions are replaced by their on-shell values, then $D_\mathit{zm}(t)$ takes the form of an oscillation with decaying amplitude. This is the standard manifestation of plasmon damping \cite{therm}, where the plasmon damping rate is $\gamma=\Pi_i(m)/2m$.

We thus determine the plasmon decay rate $\gamma$ by fitting the resulting decaying form of $D_\mathit{zm}(t)$ to what emerges from the simulation.  We compare $\gamma$ for the Gaussian and fixed amplitude cases for a range of $ma$. For $(ma)^{-1} \lesssim5$ we find that $\gamma$'s are comparable. For larger $(ma)^{-1}$ the two $\gamma$'s rapidly depart from each other, and for $(ma)^{-1}=(15, 23, 30)$  we find that the $\gamma$ in the fixed amplitude case is $\approx (6, 28, 45)$ times \textit{smaller} than for the Gaussian case.\footnote{We also checked that analogous behavior occurs on a larger lattice, $N=512$.} This is for small coupling where the leading 2-loop contribution should dominate.\footnote{These ratios are obtained by considering different couplings in the two cases that give comparable $\gamma$'s, and then accounting for the different couplings by multiplying the ratio of these $\gamma$'s by the inverse ratio of squared couplings.} Thus we once again see that the choice of ensemble has a significant affect at the 2-loop level. Given that $\gamma$ should vanish in the quantum limit, we see that the fixed amplitude ensemble is causing the lattice simulation to move towards quantum-like behavior.

We should stress that ``fixed amplitudes'' refers only to how the amplitudes are chosen in the initial configurations; after the initial time the amplitude of each mode is free to evolve as the dynamics dictates. The large reduction in the plasmon decay rate indicates that these configurations are more stable; thermalization processes are much slower than for Gaussian classical simulations. When we next study the real part of the self-energy we are forced to use the fixed amplitude ensemble, since it is the slow thermalization that makes the investigation of this quantity possible.

$\Pi_r^\textrm{2-loop}(m)$ is more difficult to extract than $\Pi_i^\textrm{2-loop}(m)$, since there is also the real 1-loop contribution. The latter is not infrared dominated (it is log divergent in the continuum) and thus completely dominates the 2-loop contribution for small coupling. To isolate the 2-loop contribution, values of $\Pi_r(m)$ may be extracted for a range of the effective dimensionless coupling $\lambda/m^2$ so that both the 1- and 2-loop contributions can be fit simultaneously to the measured linear plus quadratic dependence. This procedure was used in \cite{D}, where its relation to the 2-loop gap equation was described.

We shall adopt a complementary procedure here and will vary $\hbar$ rather than $\lambda/m^2$, since the 1-loop and 2-loop diagrams are proportional to $\hbar$ and $\hbar^2$ respectively. In the classical theory $\hbar$ originates in the normalization of the mode amplitudes, so we can just vary this overall normalization. It is a good check on the simulation to find that these two methods give similar results. As we vary $\hbar$ we also vary the bare mass $m_0$ in the field equation so as to hold the physical mass $m$ fairly constant. It is the difference between $m$ and $m_0$ that is fit to terms linear and quadratic in $\hbar$, and thus obtaining the 1- and 2-loop contributions to $\Pi_r(m)$.

For this analysis we use the 2-point function $D_t(t)$ rather than $D_\mathit{zm}(t)$. In $1+1$ dimensions the form of $D_t(t)$ reflects the contributions from both high and low momentum; it has both a rapidly oscillating component and an oscillation with $1/m$ time scale. Since $D_t(t)$ receives contributions from all the modes, it is expected to be a more reliable quantity than $D_\mathit{zm}(t)$.

We first consider the 1-loop mass correction, since the extracted value of this quantity displays a small deviation from the naively expected result, as was noted in \cite{D}. Indeed this discrepancy arises from the nonlocal correlations, as we have already described in the previous section as a finite volume effect. The effect increases for smaller $ma$, and amounts at most to about a 3\% reduction in the 1-loop diagram. The observed and calculated values of this discrepancy agree, and this is a further indication that the simulation sees the nonlocal correlations. Returning to our fit of the linear and quadratic behavior, we now input our calculated linear behavior and use the data to extract the quadratic behavior only.\footnote{The fact that this was not done in \cite{D} accounts for the slightly different results there.}

\begin{figure}
\begin{center}\includegraphics[%
  scale=0.6]{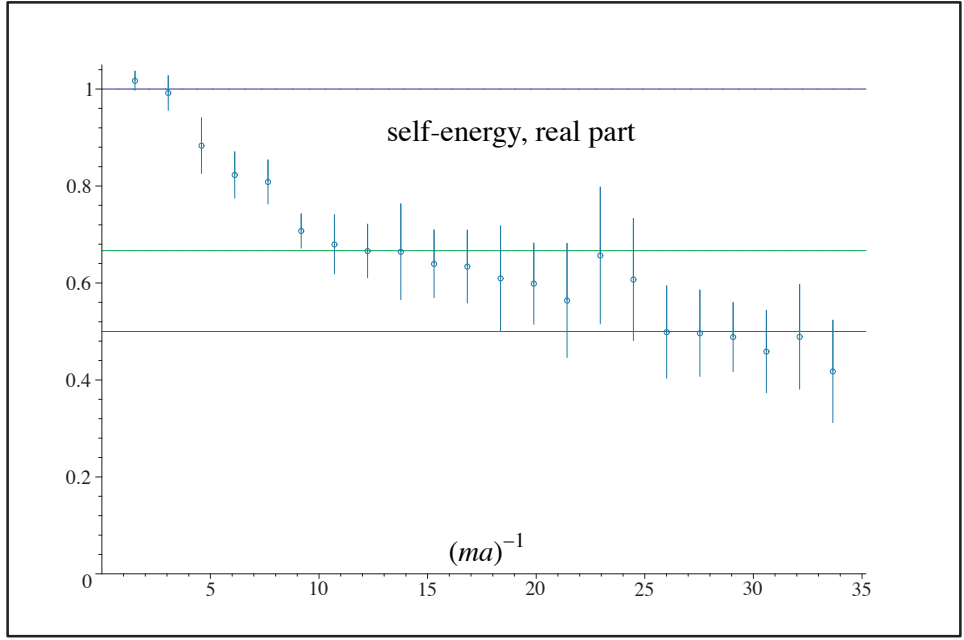}\end{center}
  \vspace{-3ex}
\caption{The real part of the 2-loop self-energy, $\Pi_r^\textrm{2-loop}(m)$, extracted from the simulation and divided by the Gaussian classical result. The quantum result at $2/3$ and the fixed amplitude result at $1/2$ are also indicated. We set $\lambda/m^2=0.4$ and let $\hbar$ be .25, .5, .75 and 1. We evolve 5000 configurations for each value of $\hbar$ and then perform a fit. We repeat this 10 times to obtain an estimate of the error, for each value of $ma$.\label{A} }
\end{figure}
\begin{figure}
\begin{center}\includegraphics[%
  scale=0.6]{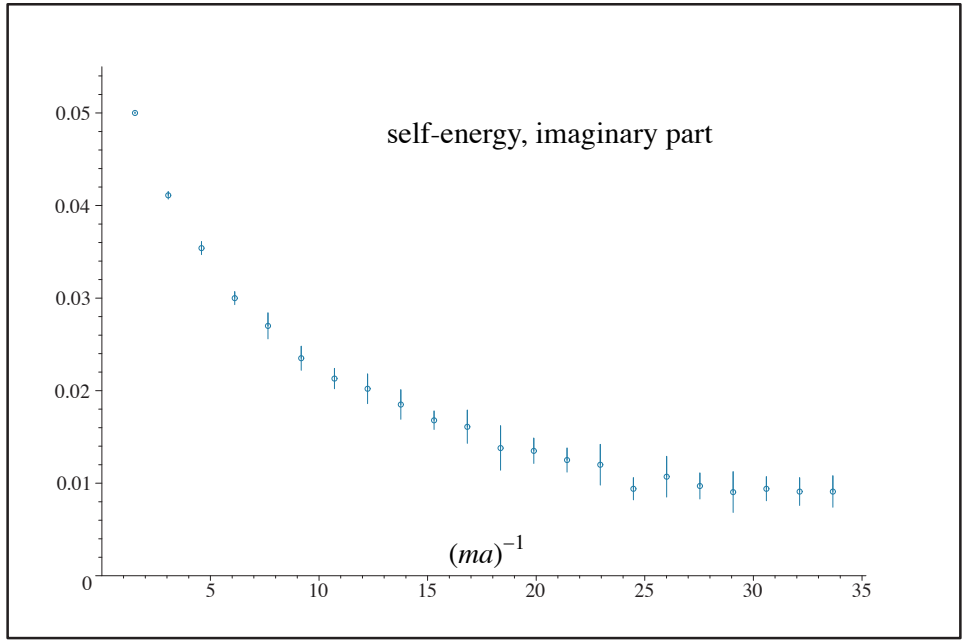}\end{center}
  \vspace{-3ex}
\caption{The imaginary part, $\Pi_i^\textrm{2-loop}(m)/m^2$, extracted from the simulation. Unlike the previous figure, only $\hbar=1$ is used.\label{B} }
\end{figure}
We compare the simulations to the explicit calculations of classical (Gaussian with $\alpha=0$) and quantum ($\alpha=1$) values of $\Pi_r^\textrm{2-loop}/m^2$, which at $\lambda/m^2=0.4$ are 0.022 and 0.015 respectively. In Fig.~(\ref{A}) we display the extracted values of $\Pi_r^\textrm{2-loop}$ divided by the classical value as a function of $(m a)^{-1}$. We see that as $ma$ decreases that $\Pi_r^\textrm{2-loop}$ gradually drops below the quantum value and becomes consistent with the calculated value of $1/2$ from the previous section.

In Fig.~(\ref{B}) we display the associated values of $\Pi_i^\textrm{2-loop}/m^2$, which was fit simultaneously to the same simulation data. In the context of the plasmon decay rate we have already noted how small this quantity is compared to its Gaussian classical value. In the latter case this quantity increases with increasing $(ma)^{-1}$, but with the fixed amplitude prescription it falls towards the vanishing quantum value.

\section{Interpretation}

We have used lattice simulations at weak coupling to extract 2-loop effects, and have confirmed that the choice of ensemble can have a significant effect, by comparing the Gaussian and fixed amplitude prescriptions. We then focused on the latter and found interesting and almost quantum-like behavior in the real and imaginary 2-loop corrections for a range of $ma$. The question now is what is the meaning of this range of $ma$.

Some insight is provided by our direct evaluation of the 2-loop mass shift $\Pi_r^\textrm{2-loop}(m)$ in section \ref{S1}. For the latter we found that we had to take the limit $\epsilon\rightarrow0$ before the $V\rightarrow\infty$ limit. If $V$ is measured with respect to $1/m$ then for the simulation with a fixed number of lattice points, increasing $V$ corresponds to increasing $m$. And $\epsilon$ corresponds loosely to the measured plasmon decay rate $\gamma$. Thus if the same order of limits is to apply in the simulation then there is an upper bound on $V$ or $m$, since $\gamma$, although small, is not exactly zero in the simulation. Thus we see why quantum-like behavior requires smaller $m$.

What is nontrivial is the fact that $\gamma$ is as small as it is, allowing nearly quantum-like behavior for a range of $ma$. Our direct evaluation in section \ref{S1} did not anticipate this, suffering as it did with a $\epsilon\rightarrow0$ divergence. The real part of the self-energy was properly anticipated, but this basically \textit{assumed} that $\gamma$ would turn out to be sufficiently small.

We note that the interesting behavior is occurring for values of $(ma)^{-1}$ that are smaller than where conventional finite volume effects are expected, which is when $(ma)^{-1}$ becomes closer to $N$. In fact finite volume effects do appear to show up in the simulation for $(ma)^{-1}\gtrsim 35$. There the mass extracted after long evolution times shows a tendency to drift downward. A drift down would result in spuriously large values of $\Pi_r^\textrm{2-loop}(m)$, and thus these finite volume effects behave oppositely to the observed departure of $\Pi_r^\textrm{2-loop}(m)$ from the Gaussian classical value.

To more directly test for conventional finite volume and discretization effects we performed a direct lattice calculation of the 2-loop graph from quantum field theory, after Wick rotation \cite{D}. The loop integrations become sums over the discrete Euclidean momenta on a $N\times N$ lattice.
\begin{eqnarray}
\Pi^\mathrm{2\;loop\;latt}(k) & = & \frac{1}{N^{4}}\sum_{\{p_{1},p_2,q_1,q_2\}=-N/2}^{N/2-1}G(p_{1}+k,p_{2})G(q_{1},q_{2})G(p_{1}+q_{1},p_{2}+q_{2}),\\
G(p_{1},p_{2}) & = & \frac{1}{4\sin^2(\pi p_{1}/N)+4\sin^2(\pi p_{2}/N)+m^{2}}\end{eqnarray}
This yields the self-energy for space-like or vanishing momentum. Of interest is the $m$ dependence of the result, and we find for $N=256$ that $\Pi_r^\textrm{2-loop-latt}(0)/m^2$ varies by less than 3\% for $10<(ma)^{-1}<48$. We may also compare this lattice result to a non-Wick-rotated calculation of the same diagram in the continuum limit, along the lines of section \ref{S1} but evaluated at zero momentum. We find agreement to within a few percent in the same range of $ma$. Thus conventional discretization or finite volume effects are significantly smaller than the effects induced by the choice of ensemble.\footnote{Discretization could be affecting the results for $(ma)^{-1}\lesssim10$, but this is not the main region of interest in this work.}

We mention a few more details of the simulation. $D_t(t)$ is measured over a time interval equal to half the spatial size of the lattice $\Delta t=a N/2$. We evolved each configuration forward for a total time equal to $(n_\Delta+1)\Delta t$, and then measured $D_t(t)$ on each of the $\Delta t$ time intervals except the first.\footnote{The first time interval allows some time for the free initial configurations to evolve.}$^,$\footnote{In our use of $D_\mathit{zm}(t)$ to determine the plasmon decay rate, discussed above, we measured $D_\mathit{zm}(t)$ once over the whole time interval $n_\Delta\Delta t$. Unlike $D_\mathit{zm}(t)$, $D_t(t)$ drastically changes its behavior for time intervals larger than $\approx2\Delta t$.} $n_\Delta=1$ for the largest value of $m$ and we scaled $n_\Delta$ as $1/m$ for smaller values of $m$. By averaging $D_t(t)$ over this many time intervals for each configuration we reduced the noise in the extracted masses. By using smaller $n_\Delta$ for larger $m$ we avoided the phenomena observed in \cite{C}, namely a slight upward drift in the masses observed for increasing evolution time (if the evolution time was increased much beyond $(n_\Delta+1)\Delta t$). The drift is more pronounced for larger mass. In fact we suspect that the enhanced drift for larger mass is related to the observation that the system behaves more classically for larger mass; greater plasmon damping implies enhanced thermalization processes, meaning that our field configurations are less stable for larger mass.

\section{Strong Coupling}
In this section we shall describe some evidence of quantum-like behavior occurring beyond the 2-loop level. At sufficiently strong coupling the quantum theory is known to have a broken phase where $\langle \phi \rangle\neq 0$, with a critical line in the $m_0$-$\lambda$ plane separating the two phases. This is usually defined as a line of fixed $g=\lambda/m_\mathrm{gap}^2$ where $m_\mathrm{gap}^2$ is the 1-loop gap mass. We may study this phenomena in the classical theory, where a priori there is no reason for the onset of $\langle \phi \rangle\neq 0$ to occur at a coupling $g$ similar to the quantum critical coupling.

A direct comparison of the quantum and classical lattice theories has the complication that the lattice regularizations are usually different. A square Euclidean lattice is often used for quantum Monte Carlo simulations, whereas the classical simulations are Lorentzian with the discretization in the time direction finer grained than the space direction. As described in \cite{D} there is a special line of constant $m_\mathrm{gap}^2$ in the $m_0$-$\lambda$ plane that is the same for the two regularizations; thus by comparing the location of the critical coupling on this line we may reduce the dependence on the lattice regularization. The result from the quantum lattice theory for the critical coupling is $g\approx10$ \cite{A}, which on the special line corresponds to $\lambda\approx50/N^2a^2$.

In the classical simulation at strong coupling we no longer have the luxury of long simulation times, but there is still a window where quantum behavior can set in before thermalization effects dominate. We determine the value of $\langle \phi \rangle$ by averaging $\phi$ over space and over times between $Na/2$ and $3Na/2$, and then averaging over configurations. The mass parameter in the initial configuration is chosen to be the value that defines the special line, $m_\mathrm{gap}^2\approx 5/N^2a^2$. A constant is also added to the initial configuration to provide a nonzero $\langle \phi \rangle$. It is adjusted to match the most probable value that is measured at the later times, since this value should be close to the minimum of the associated effective potential. To find the most probable value at later times we find the peak of the histogram obtained by binning the values of the field. After iterating so that the input value of $\langle \phi \rangle$ matches this most probable value, we then determine $\langle \phi \rangle$ as the actual average of the field at the later times. In the absence of a nonzero most probable value, $\langle \phi \rangle$ is taken to vanish.

\begin{figure}
\begin{center}\includegraphics[%
  scale=0.6]{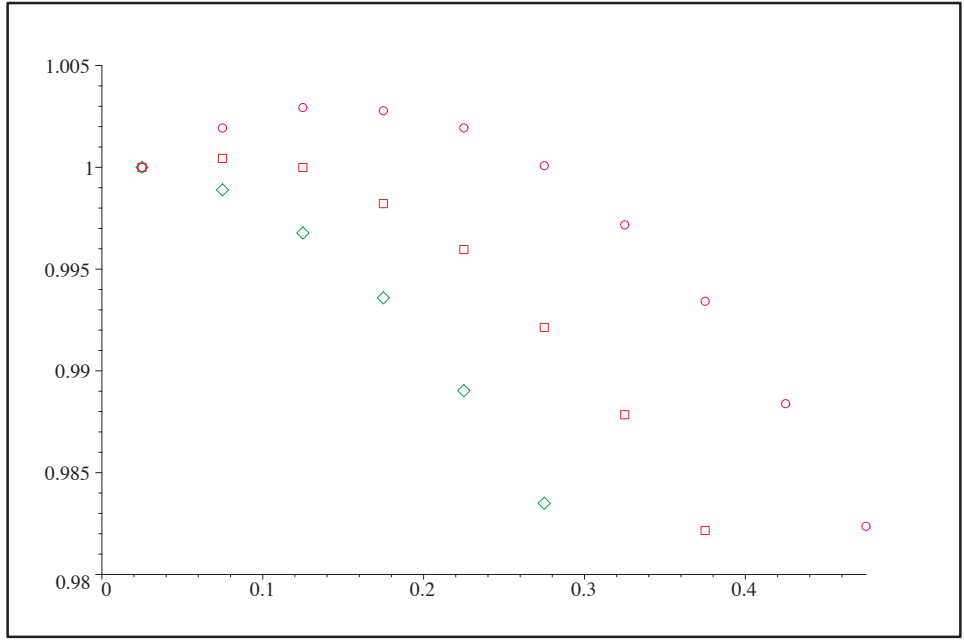}\end{center}
  \vspace{-3ex}
\caption{Histograms showing the relative probabilities of finding different values of $\langle \phi \rangle$ in the simulation, as explained in the text. The couplings are $g=(7,9,10)$ from bottom to top, the bin size is 0.05, and 10000 configurations are used for each $g$. Overall normalizations have been adjusted to equate the values in the first bin.\label{E} }
\end{figure}
We start with the fixed amplitude prescription. In Fig.~(\ref{E}) we display three examples of histograms, with $g=(7,9,10)$, with a vanishing $\langle \phi \rangle$, a slightly nonvanishing $\langle \phi \rangle$, and a larger $\langle \phi \rangle$ respectively. From this we can see that the critical coupling is a little smaller than 9. We can test for the effect of lattice size by repeating this procedure for both a smaller $N=128$ and a larger $N=512$ lattice. Here we determine the couplings that give the same slightly nonvanishing $\langle \phi \rangle$ as the $g=9$, $N=256$ case; for $N=(128, 512)$ the corresponding couplings are $g\approx (6.5, 10)$. Thus the finite size effects are diminishing for the larger lattices. We see that these results are quite consistent with a critical coupling of $g\approx 10$ as found in the quantum theory. (A less detailed analysis of this effect was described in \cite{C}. In that reference another method to locate the critical line, involving the overlap of the time and space correlators, was perhaps less precise but gave similar results.)

Now we can consider the effect of using the Gaussian distribution prescription instead, keeping everything else the same. Here we find that the critical behavior shifts towards weaker coupling; the coupling that gives the same slightly nonvanishing $\langle \phi \rangle$ is now $g\approx3$ or 4. The clearly discernible difference between these two cases again shows the effect of the choice of ensemble, this time at strong coupling. And once again it is the fixed amplitude prescription that produces quantum-like behavior.

\section{Comments}

1) In section 1 we noted that the amount of nonlocal correlations in ensembles of classical fields was controlled by the probability distribution for the amplitudes. For the 4-point function we parametrized the departure from locality by $\beta$, where a Gaussian distribution has $\beta=0$. In section 3 we described a class of Gaussian theories having a new interaction term with coefficient $\alpha$, for which $\alpha=1$ corresponded to quantum field theory. Thus the first class of theories is nonlocal and the second is nonclassical, and they could be thought of as different ways to deform the $\alpha=\beta=0$ theory which is both local and classical. All these theories have $\hbar$ and a loop expansion. In fact the two classes of theories belong to a larger class of theories in which both nonlocal and nonclassical features are introduced in varying amounts (both $\alpha$ and $\beta$ nonvanishing). But in this work we have focused on nonlocal effects in interacting classical theories and have shown that they can resemble the nonclassical effects of the corresponding local quantum field theories. These two classes of theories may have more in common than commonly thought.

2) We observe that the existence of spatial dimensions are crucial for producing quantum-like behavior. In $0+1$ dimensions then there is just one site and one mode. Averaging over the phase and holding the amplitude fixed in this case gives results that are in no way quantum-like. In fact the 1-loop correction to the mass is $1/2$ its quantum-mechanical value, as can be deduced from our discussion of the correction to the 1-loop contribution in section \ref{S1}. This can also be found through a direct perturbative solution of the $0+1$ dimensional field equation. Thus as the number of lattice points in the spatial direction becomes smaller, we would expect our results to move away from quantum-like behavior, and approach the non-quantum-mechanical behavior of the $0+1$ dimension limit. But in the other limit of fine-grained spatial dimension(s), a nonrelativistic limit of the theory could presumably be taken as done in quantum field theory, and in this way making an approach to something that has some resemblance to quantum mechanics \cite{M}.

3) The stability of the classical system and quantum-like behavior are both related to the vanishing of the imaginary self-energy on mass shell. We found some tendency for this vanishing in the lattice simulation even though there was a $\epsilon\rightarrow0$ divergence in this quantity in perturbation theory. Can the perturbative description be improved, possibly in some other classical theory?

4) The 2-loop sunset diagram turned out to be maximally sensitive to the nonlocal correlations of classical fields. How general is that result?

5) In the examples we have given, the quantum-like behavior has emerged in quantities that are infrared dominated. Is it essential that the theory itself be infrared dominated, and do asymptotically free gauge theories qualify?

\section*{Acknowledgments}
I thank T. Hirayama for discussions and S. Jeon for patiently explaining the Schwinger-Keldysh formalism to me. This work was supported in part by the National Science and Engineering Research Council of Canada.

\end{document}